\documentstyle[12pt]{article}
\newcommand{\Z}{{\sf Z \!\!\! Z}}

\makeatletter
\@addtoreset{equation}{section}
\makeatother
\renewcommand{\theequation}{\thesection.\arabic{equation}}
\def\dateandnumber(#1)#2#3#4{
\vbox to 18mm{%
     \hbox to \textwidth{ \hspace*{14mm} \hsize=40mm%
            \vbox{%
                 \hbox to 40mm{\large #1 \hss}%
                 \hbox to 40mm{    \hss}%
                 \hbox to 40mm{    \hss}%
                 }%
                 \hss \hsize=80mm%
            \vbox{%
                 \hbox to 80mm{\hss \large #2}
                 \hbox to 80mm{\hss \large #3}
                 \hbox to 80mm{\hss \large #4}
                 }%
            \hspace*{14mm} }%
      \vss
    }
}
\def\titleofpreprint#1#2#3#4{{\LARGE \bf
\vbox to 43mm{%
     \vss
     \hbox to \textwidth{ \hspace*{14mm} \hsize=130mm%
            \hss \vbox{
                      \hbox to 130mm{\hss \LARGE \bf #1\hss}%
                      \hbox to 130mm{\hss \LARGE \bf #2\hss}%
                      \hbox to 130mm{\hss \LARGE \bf #3\hss}%
                      \hbox to 130mm{\hss \LARGE \bf #4\hss}%
                 }%
            \hss \hspace*{14mm} }%
      \vss
    }}
}
\def\listofauthors#1#2#3{{\large
\vbox to 22mm{%
     \vss
     \hbox to \textwidth{ \hspace*{14mm} \hsize=130mm%
            \hss \vbox{
                      \hbox to 130mm{\hss \large #1\hss}%
                      \hbox to 130mm{\hss \large #2\hss}%
                      \hbox to 130mm{\hss \large #3\hss}%
                 }%
            \hss \hspace*{14mm} }%
      \vss
    }}
}
\def\listofaddresses#1#2#3#4#5{{\small
\vbox to 18mm{%
     \vss
     \hbox to \textwidth{ \hspace*{14mm} \hsize=130mm%
            \hss \vbox{
                      \hbox to 130mm{\hss \small #1\hss}%
                      \hbox to 130mm{\hss \small #2\hss}%
                      \hbox to 130mm{\hss \small #3\hss}%
                      \hbox to 130mm{\hss \small #4\hss}%
                      \hbox to 130mm{\hss \small #5\hss}%
                 }%
            \hss \hspace*{14mm} }%
      \vss
    }}
}
\def\abstractofpreprint#1{{\normalsize
  \vbox to 110mm{%
     \vss
     \hbox to \textwidth{\hss \normalsize \bf Abstract \hss}%
     \vspace*{1cm} \normalsize
     #1
     \vss
    }}
}

\def\footnoteitem(#1)#2{
\begin{list}{#1}{\labelwidth4.0mm \leftmargin7.0mm
\labelsep2.5mm \rightmargin7.0mm \parsep0.5ex plus0.2ex minus0.1ex
\itemsep0ex plus0.2ex }
\item #2
\end{list}
}
\begin{document}
\dateandnumber(July 1992)%
{HLRZ J\"ulich 92-47 }%
{Bern BUTP-92/34     }%
{                    }%
\titleofpreprint%
{      The confined-deconfined Interface                       }%
{      Tension and the                                         }%
{      Spectrum of the Transfer Matrix                         }%
{                                                              }%
\listofauthors%
{B. Grossmann$^1$, M. L. Laursen$^1$, T. Trappenberg$^{1,2,*}$ }%
{ and U.-J. Wiese$^{3,\#}$                                     }%
{                                                              }%
\listofaddresses%
{\em $^1$HLRZ c/o KFA J\"ulich, P.O.Box 1913, 5170 J\"ulich, Germany}%
{\em $^2$Institute of Theoretical Physics E, RWTH Aachen,}%
{\em Sommerfeldstr., 5100 Aachen, Germany }%
{\em $^3$ Universit\"at Bern, Sidlerstrasse 5, 3012 Bern, Switzerland}%
{\em                                                                 }%
\abstractofpreprint{
The reduced tension $\sigma_{cd}$
of the interface between the confined and the deconfined phase
of $SU(3)$ pure gauge theory is related to the finite size effects
of the first transfer matrix eigenvalues.
A lattice simulation of the transfer matrix spectrum
at the critical temperature
$T_c = 1/L_t$ yields $\sigma_{cd} = 0.139(4) T_c^2$ for $L_t = 2$.
We found numerical evidence
that the deconfined-deconfined domain walls
are completely wet by the confined phase, and that the
confined-deconfined interfaces are rough.
}
\begin{flushleft}
$*${Supported by the Deutsche Forschungsgemeinschaft.} \\
$\#${Supported by the Schweizer Nationalfond.} \\
\end{flushleft}
\pagebreak
\section{Introduction}

At high temperatures $QCD$ does not confine color and chiral symmetry is not
spontaneously broken. In the early universe, at about $10^{-3}$ seconds after
the big bang, quarks and gluons have passed a transition to the low temperature
confined phase in which chiral symmetry is spontaneously broken.
Depending on the quark masses the transition may be a first or second order
phase transition or just
a cross over. At a first order phase transition the nucleation of
bubbles of the confined phase from the supercooled
high temperature quark-gluon
plasma produces spatial inhomogeneities in the baryon density. If such
inhomogeneities are large enough, they may influence the primordial
nucleosynthesis of light elements which is completed
within the first   minute after the big bang. Small fluctuations,
  on the other hand, are washed
out and the baryon density at the time of nucleosynthesis becomes homogeneous.
The size of the inhomogeneities is governed by
the value of the interface tension between the high and the low temperature
phase. It is convenient to introduce the reduced interface tension
\begin{equation}
\sigma_{cd} = \frac{F}{A T},
\end{equation}
where $F$ is the interface free energy, $A$ is the interface area and $T$ is
the temperature. Computing $\sigma_{cd}$ is a nonperturbative problem which is
numerically very difficult when quarks are present.
Light quarks have a tendency to weaken the phase transition.
To obtain an upper limit for the interface tension it is therefore sufficient
to neglect quarks and to restrict oneself to a pure $SU(3)$ gauge theory of
only gluons. Then, as numerical lattice simulations have shown \cite{Gav89},
the
phase transition is first order. The value of the reduced interface
tension has been determined numerically at the critical temperature
$T_c = 1/L_t$ where $L_t$ is the
lattice extent in the euclidean time direction. For $L_t = 2$ the Boston
group obtained $\sigma_{cd} = 0.12(2) T_c^2$ \cite{Hua90} while the Helsinki
group quotes $\sigma_{cd} = 0.08(2) T_c^2$ \cite{Kar90}. In
this paper, using a completely different method, we obtain
$\sigma_{cd} = 0.139(4) T_c^2$. Closer to the
continuum limit for $L_t = 4$ the Boston group quotes
$\sigma_{cd} = 0.027(4) T_c^2$ \cite{Bro92} .
Taking this value as an
estimate of the interface tension in real QCD with quarks the typical
scale of inhomogeneities (e.g. the distance between nucleation centers)
is at most
a few centimeters \cite{Ban92}. The proton diffusion length, on the other hand,
is much larger (about 50 cm) such that the fluctuations are washed out before
primordial nucleosynthesis takes place.

In previous numerical studies of the interface tension coexistence of the
confined and the deconfined phase was enforced by keeping different parts
of the lattice at different temperatures or by applying an external field. This
breaks translation invariance and pins the interface at a certain position.
The properties of a pinned interface will in general be different
from the ones of the free interface one is interested in. To
extract the interface tension one should first perform the infinite volume
limit
and then turn off the temperature gradient or the external field. In practice
this is difficult because in a numerical simulation the lattice size
is necessarily limited. This causes finite size effects which must be
understood before one can extrapolate results reliably to the infinite volume
limit. In this work we use a different strategy. We make use of the fact that
the finite volume alone supports coexisting bulk phases when a cylindrical
geometry is used. The lattices we use have two rather short $x$- and
$y$-directions and one much longer $z$-direction. Then, without artificial
temperature gradients or external fields, several bulk phases, aligned along
the $z$-direction,
coexist with each other, separated by confined-deconfined interfaces spanned
in the short $x$- and $y$-directions. The presence of the interfaces influences
the spectrum of the transfer matrix in the long $z$-direction. In particular,
the lowest energy levels show a characteristic dependence on the interface area
and on the reduced interface tension $\sigma_{cd}$. We will use
this finite size effect to extract the value of $\sigma_{cd}$. This method
has been applied before to the four-dimensional \cite{Jan88} and to
the three-dimensional
Ising model \cite{Mey90}. It has the advantages that no pinning of the
interfaces is necessary and that finite size
effects are well understood such that the results of numerical simulations
can be extrapolated reliably to the infinite volume limit.

For the pure glue theory, however, additional phenomena arise. Since in
the high temperature phase the $\Z(3)$ center symmetry of the nonabelian gauge
group is spontaneously broken there are actually three deconfined phases. At
temperatures above $T_c$ these phases coexist. They are
separated by deconfined-deconfined
domain walls with another reduced interface tension $\sigma_{dd}(T)$
which depends on the temperature. At very high temperatures ($T \gg
\Lambda_{QCD}$) the reduced interface tension has been computed semiclassically
\cite{Bha91}
\begin{equation}
\sigma_{dd}(T) = \frac{8 \pi^2}{9 g^2} T^2,
\end{equation}
where $g$ is the gauge coupling renormalized at the scale $T$.
Frei and Patk\'{o}s \cite{Fre89}
have suggested that
\begin{equation}
\sigma_{dd}(T_c) = 2 \sigma_{cd},
\end{equation}
i.e. at the transition temperature
a deconfined-deconfined domain wall costs the same free energy as two
confined-deconfined interfaces. In this situation --- called complete wetting
--- the phase transition proceeds via domain wall splitting. Already slightly
above the transition temperature the deconfined-deconfined domain walls
split into two confined-deconfined interfaces by creating a complete wetting
layer of confined phase between two bulk deconfined phases \cite{Jue92}.
Complete wetting
is a critical phenomenon of interfaces which arises although the bulk phase
transition is first order and not universal. The critical exponents of
complete wetting were determined in ref.\cite{Tra92}. In particular, the
thickness of the confined wetting layer grows to a macroscopic size as
\begin{equation}
z_0 \propto - \log(T - T_c),
\end{equation}
and the expectation value of the Polyakov loop at the interface vanishes as
\begin{equation}
\Phi_1(0) \propto - \sqrt{T - T_c}.
\end{equation}
Numerical evidence for complete wetting has been reported for the $SU(3)$
gauge theory \cite{Kar90,Kaj91} and for the three state Potts model
\cite{Kar91}.
Complete wetting causes characteristic finite volume effects in the spectrum
of the transfer matrix \cite{Tra92}. They can be used to identify complete
wetting in a numerical simulation and to determine the value of the reduced
interface tension $\sigma_{cd}$. When wetting is incomplete, i.e.
when $\sigma_{dd}(T_c) < 2 \sigma_{cd}$, the transfer matrix spectrum is
different but
again allows to determine the values of the interface tensions \cite{Tra92}.
Wetting is a phenomenon   which arises only in the pure glue system. In the
real world with quarks the $\Z(3)$ center symmetry is explicitly broken.
Therefore, two of the three deconfined phases are only metastable. In the early
universe the metastable phases have converted into the stable one already
$10^{-14}$ seconds after the big bang \cite{Ign92}. Consequently,
deconfined-deconfined interfaces did not survive until the QCD phase transition
and wetting could not arise.
To avoid complications due to wetting in numerical simulations of the pure
glue system one may choose $C$-periodic boundary
conditions \cite{Pol91}. Just like quarks they break the
$\Z(3)$ center symmetry explicitly and eliminate two of the three
deconfined phases \cite{Wie92}. Therefore, in a $C$-periodic volume the
confined phase coexists with only one deconfined phase such  that
deconfined-deconfined domain walls are absent.

In order to apply the     finite size formulae derived in
 ref.\cite{Tra92} one has to be at the finite
volume critical temperature, i.e. at the temperature where the
free energies of the confined and the deconfined phases are equal. In practice
this temperature is not known with high enough precision. Therefore,
in this paper
we generalize the finite size formulae to temperatures slightly off $T_c$ using
a dilute interface approximation. We distinguish the cases of complete
and incomplete wetting and we also consider systems with $C$-periodic boundary
conditions. For a wide class of models similar
formulae have been derived rigorously by Borgs and Imbrie \cite{Bor91} for
rigid interfaces.
The case of the Potts model has been worked out in great detail
by Borgs \cite{Bor92}. Wang and DeTar \cite{Wan92} applied similar
formulae to results of numerical simulations of the three state Potts
model.

In three dimensions interfaces living on a lattice have a
roughening transition \cite{Bri86}. For $SU(3)$ pure lattice gauge theory
with a fixed number of lattice points in the euclidean time direction we
expect that the deconfined-deconfined interfaces become rigid  deep in the
deconfined phase, i.e.
when the Wilson coupling $\beta$ becomes large.
A rigid interface is more or less flat with only
small steplike excitations following the lattice structure. Rigid interfaces
are therefore lattice artifacts. (Of course, in condensed matter systems the
lattice often has physical reality. Then rigid
interfaces can exist in nature.) Below the roughening
transition, which we believe to be at
$\beta$-values well above the deconfinement
phase transition, the interfaces fluctuate more freely.
In particular, there exist soft modes --- the so-called
capillary waves --- which dominate the interface fluctuations
\cite{Bri86}. We expect that
confined-deconfined interfaces are always rough. Of course, at any given
physical temperature also
the deconfined-deconfined interfaces should become rough and hence forget about
the lattice structure  in the continuum limit. The continuum limit is also
taken
at large $\beta$, but now the temperature remains fixed in physical units
of e.g. the inverse bulk correlation length.
The finite size formulae we are using to extract the confined-deconfined
reduced interface tension are valid both for rough and for rigid interfaces.
Still, a comparison with the results of Borgs and Imbrie \cite{Bor91}
allows to distinguish between the two possibilities.

The paper is organized as follows. In section 2 the $\Z(3)$-symmetry
and the behavior of the
Polyakov loop are discussed in periodic and in $C$-periodic volumes.
In section 3 the finite size effects of
the spectrum of the transfer matrix are derived for complete and for
incomplete wetting as well as in the case of $C$-periodic boundary conditions.
Section~4 contains the results of our numerical simulations on the
subject of wetting and the determination of the
reduced confined-deconfined interface tension $\sigma_{cd}$.
Finally, in section~5 we summarize our results and we draw some conclusions.
In the appendix we give some details of the formulae used in section~3
to derive the spectrum in the vicinity of the phase transition.

\section{Polyakov loops in periodic and in $C$-periodic volumes}

Let us consider an $SU(3)$ pure gauge theory on an $L_x \times L_y \times
L_z \times L_t$ lattice corresponding to
a  temperature $T = 1/L_t$. The link variables $U_\mu(\vec{x},t)
\in SU(3)$ are periodic in the euclidean time direction
\begin{equation}
U_\mu(\vec{x},t + L_t) = U_\mu(\vec{x},t).
\end{equation}
The order parameter for the
            deconfinement phase transition is the Polyakov loop
\begin{equation}
\Phi(\vec{x}) = \mbox{Tr} \prod_{t = 1}^{L_t} U_4(\vec{x},t).
\end{equation}
$\Phi(\vec{x}) = \Phi_1(\vec{x}) + i \Phi_2(\vec{x})$
is a complex scalar field in three dimensions. The Polyakov loop is
invariant under periodic gauge transformations $g(\vec{x},t) \in SU(3)$.
However, under transformations
\begin{equation}
g(\vec{x},t + L_t) = g(\vec{x},t) z,
\label{gauge}
\end{equation}
which are periodic only up to a center element $z \in \Z(3) = $
$\{\exp(2\pi i n/3), n = 1,2,3\}$ it transforms as
\begin{equation}
\Phi(\vec{x})' = \Phi(\vec{x}) z,
\end{equation}
while the (Wilson) action remains invariant. Hence, the Polyakov loop indicates
if the $\Z(3)$ center symmetry is spontaneously
broken. This is the case at high temperatures.
Then color is deconfined and an external static quark has a finite free energy
$F$ such that $\langle \Phi \rangle \propto \exp(- F/T) \neq 0$. At low
temperatures, because of confinement, the free energy diverges such that
$\langle \Phi \rangle = 0$ and the $\Z(3)$-symmetry is unbroken. The two
phases are separated by a first order phase transition \cite{Gav89}. At the
transition
temperature the confined phase coexists with three degenerate deconfined
phases which are distinguished by the expectation value of the Polyakov loop.
There is one phase with a real value $\langle \Phi \rangle = \Phi^{(1)} =
(\Phi_0,0)$. The other two deconfined phases are $\Z(3)$ copies of it with
$\langle \Phi \rangle = \Phi^{(2)} = (- \frac{1}{2} \Phi_0,\frac{\sqrt{3}}{2}
\Phi_0)$ and with
$\langle \Phi \rangle = \Phi^{(3)} = (- \frac{1}{2} \Phi_0,- \frac{\sqrt{3}}{2}
\Phi_0)$.

In the numerical simulations we use lattices with a cylindrical geometry,
i.e. with two short $x$- and $y$-directions and one much longer $z$-direction.
Then close to the critical temperature typical configurations consist of
several coexisting bulk phases, aligned along the $z$-direction, and separated
by interfaces which are spanned in the transverse $x$- and $y$-directions.
In all calculations we use $C$-periodic boundary conditions \cite{Pol91} in the
$z$-direction, i.e.
\begin{equation}
U_\mu(\vec{x} + L_z \vec{e}_z,t) = \, ^CU_\mu(\vec{x},t) = U_\mu^*(\vec{x},t).
\label{Cperiodic}
\end{equation}
A $C$-periodic field is replaced by its charge conjugate when it is shifted
over the boundary. Eq.(\ref{Cperiodic})
implies $C$-periodic boundary conditions also for the Polyakov loop, i.e.
\begin{equation}
\Phi(\vec{x} + L_z \vec{e}_z) = \Phi^*(\vec{x}).
\end{equation}
This allows the existence of any number of deconfined-deconfined interfaces,
while with periodic boundary conditions the number of interfaces is
necessarily even.
One can also use $C$-periodic boundary conditions in the $x$- and
$y$-directions, i.e.
\begin{equation}
U_\mu(\vec{x} + L_x \vec{e}_x,t) = U_\mu (\vec{x} + L_y \vec{e}_y,t) =
U_\mu^*(\vec{x},t).
\end{equation}
The $C$-periodic action is invariant under $C$-periodic gauge transformations
\begin{equation}
g(\vec{x} + L_i \vec{e}_i,t) = g^*(\vec{x},t).
\end{equation}
Considering again a transformation which is up to a center element periodic
in the euclidean time direction one finds
\begin{equation}
g^*(\vec{x},t) = g^*(\vec{x},t + L_t) z = g(\vec{x} + L_i \vec{e}_i,t + L_t) z
=
g(\vec{x} + L_i \vec{e}_i,t) z^2 = g^*(\vec{x},t) z^2.
\end{equation}
Consistency requires $z^2 = 1$ and hence $z = 1$ (because $z \in \Z(3)$). The
transformations of eq.(\ref{gauge}) are inconsistent with $C$-periodic
boundary conditions unless $z = 1$. Hence, in a $C$-periodic volume the
$\Z(3)$-symmetry is explicitly broken by the boundary conditions.
With $C$-periodic
boundary conditions in the $x$- and $y$-directions the Polyakov loop obeys
\begin{equation}
\Phi(\vec{x} + L_x \vec{e}_x) = \Phi(\vec{x} + L_y \vec{e}_y) =
\Phi^*(\vec{x}).
\end{equation}
In particular, the two deconfined phases $\Phi^{(2)}$ and $\Phi^{(3)}$
with non-real expectation values
of the Polyakov loop are excluded by the boundary conditions. Hence,
in a $C$-periodic volume the confined phase
coexists only with the deconfined phase  $\Phi^{(1)}$ \cite{Wie92}.
This excludes
complications due to wetting because deconfined-deconfined domain walls
cannot even exist. On the other hand, when periodic boundary conditions are
used in the short $x$- and $y$-directions all three deconfined phases
coexist with the confined phase. In this situation one can study questions
of wetting.

In the simulations we measure correlation functions
$\langle {\cal O}_i(0) {\cal O}_i(z) \rangle$ in the long $z$-direction.
The operators we use are the real and imaginary parts as well as the absolute
value of the Polyakov loop
\begin{eqnarray}
{\cal O}_1(z) = \sum_{x,y} \Phi_1(x,y,z), \nonumber \\
{\cal O}_2(z) = \sum_{x,y} \Phi_2(x,y,z), \nonumber \\
{\cal O}_3(z) = \sum_{x,y} |\Phi(x,y,z)|^2.
\label{operators}
\end{eqnarray}
The operators ${\cal O}_1$ and ${\cal O}_3$ are even under charge
conjugation and their correlation functions are therefore periodic. The
operator ${\cal O}_2$, on the other hand, is $C$-odd and its correlation
function is hence anti-periodic. A $C$-odd correlation function is easier to
measure in a numerical simulation because it gets contributions from single
deconfined-deconfined
interfaces while a nontrivial contribution to a $C$-even correlation function
requires at least two interfaces and is therefore more suppressed.

\section{The interface tension and the spectrum of the transfer matrix}

The transfer matrix ${\cal T}(z)$ describes the evolution of the gluonic
system over a distance $z$ in the spatial $z$-direction. In field
representation
it can be defined as the transition amplitude between an initial Polyakov loop
distribution $\Phi^{(i)}(x,y)$ and a final distribution $\Phi^{(j)}(x,y)$,
i.e. as a path integral
\begin{equation}
{\cal T}_{ij}(z) = \int {\cal D} U \exp(- S[U]),
\end{equation}
where $S[U]$ is the gluon action. The integration is restricted to field
configurations in the interval $[0,z]$ for which $\Phi(x,y,0) =
\Phi^{(i)}(x,y)$ and $\Phi(x,y,z) = \Phi^{(j)}(x,y)$. The transfer matrix can
be written as
\begin{equation}
{\cal T}(z) = \exp(- H z).
\label{trans}
\end{equation}
The spectrum of $H$ shows characteristic finite size effects which allow
to distinguish between complete and incomplete wetting and which can be used
to determine the value of the reduced confined-deconfined interface tension.

For simplicity let us start with a system with $C$-periodic boundary
conditions in the short $x$- and $y$-directions. Then
only the deconfined phase $\Phi^{(1)}$ coexists with the confined phase.
In the subspace of these two phases the transfer matrix takes the form
\begin{eqnarray}
t(z) = \left( \begin{array}{cc}
t_{dd}(z) & t_{cd}(z) \\
t_{cd}(z) & t_{cc}(z) \end{array} \right),
\end{eqnarray}
where $t_{dd}(z)$ is the transition amplitude from the deconfined phase
back to itself, $t_{cd}(z)$ describes transitions from the confined
phase to the
deconfined phase (or vice versa) and $t_{cc}(z)$ is the amplitude for
transitions from the confined phase back to the confined phase.
The eigenvalues of the transfer matrix are
\begin{eqnarray}
t_0(z) = \frac{1}{2}[t_{dd}(z) + t_{cc}(z) + \sqrt{(t_{dd}(z) - t_{cc}(z))^2 +
4
t_{cd}(z)^2}],
\nonumber \\
t_1(z) = \frac{1}{2}[t_{dd}(z) + t_{cc}(z) - \sqrt{(t_{dd}(z) - t_{cc}(z))^2 +
4
t_{cd}(z)^2}].
\end{eqnarray}

In ref.\cite{Tra92} the amplitudes $t_{dd}(z)$, $t_{cd}(z)$, $t_{cc}(z)$ were
computed in the dilute interface approximation assuming
that the free energies of all
phases are equal, i.e. that one is exactly at the finite volume
critical temperature. In practice, this temperature is not known with high
enough precision. Therefore, to interpret the numerical data correctly, it
is necessary to understand the transfer matrix spectrum also slightly off
$T_c$.
This requires to take the free energy difference between
the confined and the deconfined phases into account. In the dilute interface
calculation we perform a path integral over configurations which consist of
several bulk phases, aligned along the $z$-direction, and separated by
interfaces
spanned in the $x$-$y$-plane. A block of confined phase of thickness  $z_0$
costs the free energy $F = f_c A z_0$, where $f_c$ is the free energy density
of the confined phase and $A = L_x L_y$ is the area in the transverse
directions. Hence, such a block of confined phase is weighted with a
Boltzmann factor
\begin{equation}
\exp(- F/T) = \exp(- f_c A z_0 /T).
\end{equation}
Similarly, a block of deconfined phase gets a factor
\begin{equation}
\exp(- F/T) = \exp(- f_d A z_0 /T).
\end{equation}
We can  take out an overall factor
$\exp\left(- \frac{1}{2}(f_c+f_d)\, A z_0/T\right)$ without changing the energy
differences. Following ref.\cite{Bor92} we introduce the variable
\begin{equation}
x = \frac{1}{2}(f_c - f_d) A/T.
\end{equation}
Then we write a factor $\exp(-x z_0)$ for the confined phase
and a factor $\exp(x z_0)$ for the deconfined phase.
The bulk phases are separated by interfaces which cost a free energy
$F = \sigma_{cd} A T$, where $\sigma_{cd}$ is the reduced confined-deconfined
interface tension. An interface gets the Boltzmann factor
\begin{equation}
\exp(-F/T) = \delta \exp(- \sigma_{cd} A),
\end{equation}
where the pre-exponential factor $\delta$ arises due to capillary wave
fluctuations of the interface (which of course also influence the value of
$\sigma_{cd}$). In general, one would expect that $\delta$ also
depends on the interface area $A$. However, in three dimensions the leading
contribution to $\delta$ is $A$-independent \cite{Bre85}.
We treat $\delta$ as an unknown
constant. In ref.\cite{Bor91} the pre-exponential factor was computed for
rigid interfaces as $\delta = 1$.
We expect that our interfaces are rough and that
these results do not apply here.

Let us write down the leading contributions to the amplitude $t_{dd}(z)$;
\begin{eqnarray}
&&t_{dd}(z) = \exp(x z) + \int_0^z dz_0 \int_{z_0}^z dz_0' \nonumber \\
&&\exp(x z_0) \delta \exp(- \sigma_{cd} A) \exp(-x (z_0' - z_0))
\delta \exp(- \sigma_{cd} A) \exp(x (z - z_0')) + ...  \; .\nonumber \\ \,
\label{A}
\end{eqnarray}
The first term describes a block of deconfined phase of thickness $z$ filling
the whole interval. The second term is the two-interface contribution. One
integrates over the positions $z_0$ and $z_0'$ of the two interfaces. The
first factor in the integrand stands for a deconfined bulk phase of thickness
$z_0$. The next factor describes the interface located at $z_0$
followed by a block of confined bulk phase of thickness $z_0' - z_0$. Next
comes the factor for the interface located at $z_0'$ and finally a deconfined
bulk phase of thickness $z - z_0'$. Graphically one may write eq.(\ref{A}) as
\begin{equation}
t_{dd}(z) =
\begin{picture}(8,2.4)(0,0)
\put(0,2){\line(1,0){8}}
\put(0,-.4){\line(1,0){8}}
\put(4,.9){\makebox(0,0){\small $d$}}
\end{picture}
\; + \;
\begin{picture}(13,2.4)(0,0)
\put(0,2){\line(1,0){13}}
\put(0,-.4){\line(1,0){13}}
\put(4,-.4){\line(0,1){2.4}}
\put(9,-.4){\line(0,1){2.4}}
\put(2,.9){\makebox(0,0){\small $d$}}
\put(6.5,.9){\makebox(0,0){\small $c$}}
\put(11,.9){\makebox(0,0){\small $d$}}
\end{picture}
\; + \; ... \; ,
\end{equation}
and in complete analogy one constructs
the expression
\begin{equation}
t_{cc}(z) =
\begin{picture}(8,2.4)(0,0)
\put(0,2){\line(1,0){8}}
\put(0,-.4){\line(1,0){8}}
\put(4,.9){\makebox(0,0){\small $c$}}
\end{picture}
\; + \;
\begin{picture}(13,2.4)(0,0)
\put(0,2){\line(1,0){13}}
\put(0,-.4){\line(1,0){13}}
\put(4,-.4){\line(0,1){2.4}}
\put(9,-.4){\line(0,1){2.4}}
\put(2,.9){\makebox(0,0){\small $c$}}
\put(6.5,.9){\makebox(0,0){\small $d$}}
\put(11,.9){\makebox(0,0){\small $c$}}
\end{picture}
\; +  \; ... \; .
\end{equation}
In fact one gets $t_{cc}(z)$ from $t_{dd}(z)$ by replacing $x$ by $-x$.
Similarly one writes
\begin{equation}
t_{cd}(z) = \int_0^z dz_0
\exp(-x z_0) \delta \exp(- \sigma_{cd} A) \exp(x (z - z_0)) + ... =
\begin{picture}(10,2.4)(0,0)
\put(0,2){\line(1,0){10}}
\put(0,-.4){\line(1,0){10}}
\put(5,-.4){\line(0,1){2.4}}
\put(2,.9){\makebox(0,0){\small $c$}}
\put(8,.9){\makebox(0,0){\small $d$}}
\end{picture}
\; + \; ... \; ,
\end{equation}
which is the one-interface contribution to the transition amplitude between
the deconfined and the confined phase.
It is straightforward, although somewhat tedious,
to work out the multi-interface contributions and we refer the reader
to the appendix for details. Using the
variable $\delta' = \delta \exp(-\sigma_{cd}A)$ the different terms are
\begin{eqnarray}
t_{dd}(z) & = &     \cosh (z\sqrt{x^{2} + {\delta'} ^2}) +
                    \frac{x}{\sqrt{x^{2} + {\delta'} ^2}}
                            \sinh (z\sqrt{x^{2} + {\delta'} ^2}),
\nonumber\\
t_{cc}(z) & = &                 \cosh (z\sqrt{x^{2} + {\delta'} ^2}) -
                          \frac{x}{\sqrt{x^{2} + {\delta'} ^2}}
                                           \sinh (z\sqrt{x^{2} + {\delta'}
^2}),
 \nonumber \\
t_{cd}(z) & = & \frac{{\delta'}}{\sqrt{x^{2} + {\delta'}^2}}
                       \sinh (z\sqrt{x^{2} + {\delta'} ^2})
\end{eqnarray}
leading to the eigenvalues
\begin{eqnarray}
t_{0}(z) & = & \exp (z\sqrt{x^{2} + {\delta'} ^2}),\nonumber \\
t_{1}(z) & = & \exp (-z\sqrt{x^{2} + {\delta'} ^2}).
\end{eqnarray}
According to eq.(\ref{trans}) we define $E_i$ by
\begin{equation}
t_i(z) = \exp(- E_i z),
\end{equation}
and we find for the energy difference
\begin{equation}
E_1 - E_0 = 2 \sqrt{x^2 + \delta^2 \exp(- 2 \sigma_{cd} A)}.
\end{equation}
At the finite volume critical temperature (i.e. for $x = 0$) the result
agrees with ref.\cite{Tra92}. It predicts that the energy splitting
between the lowest states has a minimum at the critical temperature. At the
minimum the energy splitting
\begin{equation}
E_1 - E_0 = 2 \delta \exp(- \sigma_{cd} A) \,\,\,\,\, (\mbox{for} \, x = 0)
\label{finiteC}
\end{equation}
vanishes exponentially with the interface area times the reduced
confined-deconfined interface tension $\sigma_{cd}$. This equation will be used
to extract the value of $\sigma_{cd}$ in the numerical simulations.

Now let us turn to the more complicated case of periodic boundary conditions
in which the confined phase coexists with all three deconfined phases.
As it was shown in refs.\cite{Tra92,Bor92} in the subspace of the
three deconfined
phases $\Phi^{(1)}$, $\Phi^{(2)}$, $\Phi^{(3)}$ and the confined phase
the transfer matrix takes the form
\begin{eqnarray}
t(z) = \left( \begin{array}{cccc}
t_{dd}(z) & t_{dd'}(z) & t_{dd'}(z) & t_{cd}(z) \\
t_{dd'}(z) & t_{dd}(z) & t_{dd'}(z) & t_{cd}(z) \\
t_{dd'}(z) & t_{dd'}(z) & t_{dd}(z) & t_{cd}(z) \\
t_{cd}(z) & t_{cd}(z) & t_{cd}(z) & t_{cc}(z) \end{array} \right),
\end{eqnarray}
where $t_{dd}(z)$, $t_{dd'}(z)$, $t_{cd}(z)$ and $t_{cc}(z)$ are
transition amplitudes between the
various phases.
The transfer matrix $t(z)$ is symmetric under the permutation group
$S_3\cong \Z(3)\times C$ which is only broken by the boundary conditions in the
$z$-direction. This group has three irreducible representations, namely a
symmetric
(trivial), an antisymmetric, and a mixed symmetric one, corresponding to the
Young
tableaux
\begin{equation}
\begin{centering}
\begin{picture}(80,10.5)(0,3)
\put(10,8){\framebox(2,2)}
\put(12,8){\framebox(2,2)}
\put(14,8){\framebox(2,2)}
\put(30,4){\framebox(2,2)}
\put(30,6){\framebox(2,2)}
\put(30,8){\framebox(2,2)}
\put(50,8){\framebox(2,2)}
\put(52,8){\framebox(2,2)}
\put(50,6){\framebox(2,2)}
\end{picture}
\end{centering}
\end{equation}
It turns out that the spectrum of the above transfer matrix splits up into a
symmetric
ground state with the eigenvalue $t_{0}(z)$, a corresponding
excitation
$t_{3}(z)$, and a twofold degenerate eigenvalue $t_{1,2}(z)$ of mixed symmetry.
The eigenvectors are of the form
\begin{equation}
v_s = (v_d,v_d,v_d,v_c)
\end{equation}
for the symmetric representation, and
\begin{equation}
v_m = (v_1,v_2,v_3,0) \mbox{ with } \sum_{i=1}^{3} v_i = 0
\end{equation}
for the mixed representation. For the latter we chose one basis vector
$|1\rangle$ with positive $C$-parity and the other one $|2\rangle$ with
negative $C$-parity.
 The corresponding eigenvalues of the transfer matrix are
\begin{eqnarray}
&&t_0(z) = \frac{1}{2} [t_{dd}(z) + 2 t_{dd'}(z) + t_{cc}(z) +
\sqrt{(t_{dd}(z) + 2 t_{dd'}(z) - t_{cc}(z))^2 + 12 t_{cd}(z)^2}], \nonumber \\
&&t_{1,2}(z) = t_{dd}(z) - t_{dd'}(z), \nonumber \\
&&t_3(z) = \frac{1}{2} [t_{dd}(z) + 2 t_{dd'}(z) + t_{cc}(z) -
\sqrt{(t_{dd}(z) + 2 t_{dd'}(z) - t_{cc}(z))^2 + 12 t_{cd}(z)^2}]. \nonumber \\
\,
\end{eqnarray}
Due to the $S_3$-symmetry the eigenvalues $t_1(z)$ and $t_2(z)$ are
degenerate. We must distinguish two cases: complete and incomplete wetting.

Let us begin with the {\bf complete wetting} case. Then a deconfined-deconfined
domain wall consists of two widely separated confined-deconfined interfaces.
Therefore, in the dilute interface approximation one must include only
these interfaces, while direct deconfined-deconfined interfaces do not exist.
In graphical notation the various amplitudes read
\begin{eqnarray}
t_{dd}(z) & = &
\begin{picture}(13,2.4)(0,0)
\put(0,2){\line(1,0){13}}
\put(0,-.4){\line(1,0){13}}
\put(6.5,.9){\makebox(0,0){\small $d$}}
\end{picture}
\; + \;     \;\;
\begin{picture}(13,2.4)(0,0)
\put(0,2){\line(1,0){13}}
\put(0,-.4){\line(1,0){13}}
\put(4,-.4){\line(0,1){2.4}}
\put(9,-.4){\line(0,1){2.4}}
\put(2,.9){\makebox(0,0){\small $d$}}
\put(6.5,.9){\makebox(0,0){\small $c$}}
\put(11,.9){\makebox(0,0){\small $d$}}
\end{picture}
\; + \; ... \; , \nonumber \\
t_{dd'}(z) & = &
\begin{picture}(13,2.4)(0,0)
\put(0,2){\line(1,0){13}}
\put(0,-.4){\line(1,0){13}}
\put(4,-.4){\line(0,1){2.4}}
\put(9,-.4){\line(0,1){2.4}}
\put(2,.9){\makebox(0,0){\small $d$}}
\put(6.5,.9){\makebox(0,0){\small $c$}}
\put(11,.9){\makebox(0,0){\small $d'$}}
\end{picture}
\; + \; ... \; , \nonumber \\
t_{cd}(z) & = &
\begin{picture}(13,2.4)(0,0)
\put(0,2){\line(1,0){13}}
\put(0,-.4){\line(1,0){13}}
\put(6.5,-.4){\line(0,1){2.4}}
\put(3,.9){\makebox(0,0){\small $c$}}
\put(10,.9){\makebox(0,0){\small $d$}}
\end{picture}
\; + \; ... \; , \nonumber \\
t_{cc}(z) & = &
\begin{picture}(13,2.4)(0,0)
\put(0,2){\line(1,0){13}}
\put(0,-.4){\line(1,0){13}}
\put(6.5,.9){\makebox(0,0){\small $c$}}
\end{picture}
\; + \; 3 \;
\begin{picture}(13,2.4)(0,0)
\put(0,2){\line(1,0){13}}
\put(0,-.4){\line(1,0){13}}
\put(4,-.4){\line(0,1){2.4}}
\put(9,-.4){\line(0,1){2.4}}
\put(2,.9){\makebox(0,0){\small $c$}}
\put(6.5,.9){\makebox(0,0){\small $d$}}
\put(11,.9){\makebox(0,0){\small $c$}}
\end{picture}
\; + \; ... \; .
\end{eqnarray}
Due to complete wetting
the leading contribution to the transition amplitude $t_{dd'}(z)$
between two different deconfined phases
comes from two confined-deconfined interfaces.
The factor 3 in the amplitude $t_{cc}(z)$ arises because there are three
possible realizations of the deconfined phase between the two blocks of
confined phase. Following the appendix we now put $\delta' = \sqrt{3}
\delta \exp(- \sigma_{cd}A)$.
Working out the multi-interface contributions results in
\begin{eqnarray}
\sqrt{3}t_{cd}(z) & = & \frac{{\delta'}}{\sqrt{x^{2} + {\delta'} ^2}}
                       \sinh (z\sqrt{x^{2} + {\delta'} ^2}),\nonumber \\
t_{cc}(z) & = &              \cosh (z\sqrt{x^{2} + {\delta'} ^2}) -
                       \frac{x}{\sqrt{x^{2} + {\delta'} ^2}}
                      \sinh (z\sqrt{x^{2} + {\delta'} ^2}),\nonumber \\
t_{dd}(z) + 2t_{dd'}(z) & = &     \cosh (z\sqrt{x^{2} + {\delta'} ^2}) +
                          \frac{x}{\sqrt{x^{2} + {\delta'} ^2}}
                      \sinh (z\sqrt{x^{2} + {\delta'} ^2}),\nonumber \\
t_{dd}(z) - t_{dd'}(z) & = & \exp(zx),
\end{eqnarray}
where $t_{cc}(z)$ and $t_{dd}(z) + 2t_{dd'}(z)$
are related by exchanging $x$ with $-x$. The eigenvalues are
\begin{eqnarray}
t_{0}(z) & = & \exp (z\sqrt{x^{2} + {\delta'} ^2}),\nonumber \\
t_{1,2}(z) & = & \exp(zx), \nonumber \\
t_{3}(z) & = & \exp (-z\sqrt{x^{2} + {\delta'} ^2}).
\end{eqnarray}
resulting in the energies
\begin{eqnarray}
&&E_{1,2} - E_0 = \sqrt{x^2 + 3 \delta^2 \exp(- 2 \sigma_{cd} A)} - x,
\nonumber
\\
&&E_3 - E_0 = 2 \sqrt{x^2 + 3 \delta^2 \exp(- 2 \sigma_{cd} A)}.
\label{spectrum}
\end{eqnarray}
At the critical temperature (for $x = 0$) the results are in agreement with
ref.\cite{Tra92}. To leading order in $x$ they are identical with the finite
size formulae derived by Borgs \cite{Bor92} who neglected higher-order
multi-instanton contributions.
We note that the $x$-dependence disappears in the
combination
\begin{equation}
(E_3 - E_{1,2})(E_{1,2} - E_0) = 3 \delta^2 \exp(- 2 \sigma_{cd} A).
\end{equation}
Eq.(\ref{spectrum}) predicts that the
energy $E_3 - E_0$ has a minimum at the critical temperature ($x = 0$), at
which
\begin{eqnarray}
&&E_{1,2} - E_0 = \sqrt{3} \delta \exp(- \sigma_{cd} A), \nonumber \\
&&E_3 - E_0 = 2 \sqrt{3} \delta \exp(- \sigma_{cd} A) \,\,\,\,\,
(\mbox{for} \, x = 0),
\label{finiteP}
\end{eqnarray}
such that one can again read off the value of $\sigma_{cd}$. The equidistant
energy splitting between the lowest levels at the critical temperature, i.e.
\begin{equation}
2(E_{1,2} - E_0) = E_3 - E_0 \,\,\,\,\, (\mbox{for} \, x = 0),
\end{equation}
is characteristic for complete wetting. It disappears in the incomplete wetting
case to which we now turn.

In case of {\bf incomplete wetting} also direct deconfined-deconfined domain
walls
exist, which do not consist of two confined-deconfined interfaces. They have
another reduced interface tension $\sigma_{dd}$ and they get a
Boltzmann factor
\begin{equation}
\exp(- F/T) = \gamma \exp(- \sigma_{dd} A).
\end{equation}
The various amplitudes then get additional contributions
\begin{eqnarray}
t_{dd}(z) & = &
\begin{picture}(10,2.4)(0,0)
\put(0,2){\line(1,0){10}}
\put(0,-.4){\line(1,0){10}}
\put(5,.9){\makebox(0,0){\small $d$}}
\end{picture}
\; + \;    \;\;
\begin{picture}(13,2.4)(0,0)
\put(0,2){\line(1,0){13}}
\put(0,-.4){\line(1,0){13}}
\put(4,-.4){\line(0,1){2.4}}
\put(9,-.4){\line(0,1){2.4}}
\put(2,.9){\makebox(0,0){\small $d$}}
\put(6.5,.9){\makebox(0,0){\small $c$}}
\put(11,.9){\makebox(0,0){\small $d$}}
\end{picture}
\; + \; 2 \;
\begin{picture}(13,2.4)(0,0)
\put(0,2){\line(1,0){13}}
\put(0,-.4){\line(1,0){13}}
\put(4,-.4){\line(0,1){2.4}}
\put(9,-.4){\line(0,1){2.4}}
\put(2,.9){\makebox(0,0){\small $d$}}
\put(6.5,.9){\makebox(0,0){\small $d'$}}
\put(11,.9){\makebox(0,0){\small $d$}}
\end{picture}
\; + \; ... \; ,  \nonumber \\
t_{dd'}(z) & = &
\begin{picture}(10,2.4)(0,0)
\put(0,2){\line(1,0){10}}
\put(0,-.4){\line(1,0){10}}
\put(5,-.4){\line(0,1){2.4}}
\put(2,.9){\makebox(0,0){\small $d$}}
\put(8,.9){\makebox(0,0){\small $d'$}}
\end{picture}
\; + \;  \;\;
\begin{picture}(13,2.4)(0,0)
\put(0,2){\line(1,0){13}}
\put(0,-.4){\line(1,0){13}}
\put(4,-.4){\line(0,1){2.4}}
\put(9,-.4){\line(0,1){2.4}}
\put(2,.9){\makebox(0,0){\small $d$}}
\put(6.5,.9){\makebox(0,0){\small $c$}}
\put(11,.9){\makebox(0,0){\small $d'$}}
\end{picture}
\; + \;     \;\;
\begin{picture}(13,2.4)(0,0)
\put(0,2){\line(1,0){13}}
\put(0,-.4){\line(1,0){13}}
\put(4,-.4){\line(0,1){2.4}}
\put(9,-.4){\line(0,1){2.4}}
\put(2,.9){\makebox(0,0){\small $d$}}
\put(6.5,.9){\makebox(0,0){\small $d''$}}
\put(11,.9){\makebox(0,0){\small $d'$}}
\end{picture}
\; + \; ... \; ,  \nonumber \\
t_{cd}(z) & = &
\begin{picture}(10,2.4)(0,0)
\put(0,2){\line(1,0){10}}
\put(0,-.4){\line(1,0){10}}
\put(5,-.4){\line(0,1){2.4}}
\put(2,.9){\makebox(0,0){\small $c$}}
\put(8,.9){\makebox(0,0){\small $d$}}
\end{picture}
\; + \; 2 \;
\begin{picture}(13,2.4)(0,0)
\put(0,2){\line(1,0){13}}
\put(0,-.4){\line(1,0){13}}
\put(4,-.4){\line(0,1){2.4}}
\put(9,-.4){\line(0,1){2.4}}
\put(2,.9){\makebox(0,0){\small $c$}}
\put(6.5,.9){\makebox(0,0){\small $d'$}}
\put(11,.9){\makebox(0,0){\small $d$}}
\end{picture}
\; + \; ... \; ,  \nonumber \\
t_{cc}(z) & = &
\begin{picture}(10,2.4)(0,0)
\put(0,2){\line(1,0){10}}
\put(0,-.4){\line(1,0){10}}
\put(5,.9){\makebox(0,0){\small $c$}}
\end{picture}
\; + \; 3 \;
\begin{picture}(13,2.4)(0,0)
\put(0,2){\line(1,0){13}}
\put(0,-.4){\line(1,0){13}}
\put(4,-.4){\line(0,1){2.4}}
\put(9,-.4){\line(0,1){2.4}}
\put(2,.9){\makebox(0,0){\small $c$}}
\put(6.5,.9){\makebox(0,0){\small $d$}}
\put(11,.9){\makebox(0,0){\small $c$}}
\end{picture}
\; + \; ... \; .
\end{eqnarray}
In particular, the leading contribution to the amplitude $t_{dd'}(z)$ now comes
from a direct deconfined-deconfined interface.
Summing up all multi-interface contributions and using the
variables ${\gamma'} = 2 \gamma \exp(-\sigma_{dd}A)$ and
${\delta'} = \sqrt{3} \delta \exp(-\sigma_{cd}A)$
along with
$x' = x + {\gamma'}/2$ yields
\begin{eqnarray}
 \sqrt{3}t_{cd}(z) & = &
 \frac{{\delta'}}{\sqrt{x'^{2} + {\delta'} ^2}}
      \sinh (z\sqrt{x'^{2} + {\delta'} ^2})\exp(z{\gamma'}/2),
                   \nonumber \\
  t_{cc}(z) & = & \left(\cosh (z\sqrt{x'^{2} + {\delta'} ^2}) -
         \frac{x'}{\sqrt{x'^{2} + {\delta'} ^2}}
                     \sinh (z\sqrt{x'^{2} + {\delta'} ^2})\right)
                    \exp(z{\gamma'}/2), \nonumber \\
 t_{dd}(z) + 2t_{dd'}(z) & = &
              \left( \cosh (z\sqrt{x'^{2} + {\delta'} ^2}) +
         \frac{x'}{\sqrt{x'^{2} + {\delta'} ^2}}
                     \sinh (z\sqrt{x'^{2} + {\delta'} ^2}) \right)
                    \exp(z{\gamma'}/2), \nonumber \\
  t_{dd}(z) - t_{dd'}(z) & = & \exp(zx)\exp(-z{\gamma'}/2).
\end{eqnarray}
We arrive at the eigenvalues
\begin{eqnarray}
t_{0}(z) & = &\exp (z\sqrt{x'^{2} + {\delta'} ^2})\exp(z{\gamma'}/2)
,\nonumber \\
t_{1,2}(z) & = & \exp(zx)\exp(-z{\gamma'}/2), \nonumber \\
t_{3}(z) & = & \exp (-z\sqrt{x'^{2} + {\delta'} ^2})\exp(z{\gamma'}/2)
\end{eqnarray}
 and hence
\begin{eqnarray}
&&E_{1,2} - E_0 =
\sqrt{(x + \gamma \exp(- \sigma_{dd} A))^2 + 3 \delta^2 \exp(- 2 \sigma_{cd}
A)}
- x + 2 \gamma \exp(- \sigma_{dd} A),
\nonumber \\
&&E_3 - E_0 = 2 \sqrt{(x + \gamma \exp(- \sigma_{dd} A))^2 +
3 \delta^2 \exp(- 2 \sigma_{cd} A)}.
\end{eqnarray}
The combination
\begin{equation}
(E_3 - E_{1,2})(E_{1,2} - E_0) = 3 \delta^2 \exp(- 2 \sigma_{cd} A) -
3 \gamma^2 \exp(- 2 \sigma_{dd} A) + 6 x \gamma \exp(- \sigma_{dd} A)
\end{equation}
is now $x$-dependent. Furthermore,
the minimum of $E_3 - E_0$ no longer occurs at the critical temperature.
Instead it corresponds to
\begin{equation}
x = - \gamma \exp(- \sigma_{dd} A) < 0,
\end{equation}
which lies in the confined phase, because $x < 0$ implies that the free energy
of the confined phase is smaller than the one of the deconfined phase.
At the minimum one finds
\begin{eqnarray}
&&E_{1,2} - E_0 =
\sqrt{3} \delta \exp(- \sigma_{cd} A) + 3 \gamma \exp(- \sigma_{dd} A),
\nonumber \\
&&E_3 - E_0 = 2 \sqrt{3} \delta \exp(- \sigma_{cd} A)
\,\,\,\,\,\,\,\,\,\,\,\,\,\,\,\,\,\,(\mbox{for} \,
x = - \gamma \exp(- \sigma_{dd} A)).
\end{eqnarray}
As opposed to the complete wetting case the point of equidistant energy
splittings between the lowest levels is now at a different position
\begin{equation}
x = 2 \gamma \exp(- \sigma_{dd} A) > 0,
\end{equation}
which lies in the deconfined phase.
It is clear that the transfer matrix spectrum is qualitatively different from
the complete wetting case.

\section{Numerical results}

We have performed numerical simulations of the $SU(3)$ pure gauge theory
on lattices with $L_t = 2$ points in the euclidean time direction close
to the deconfinement phase transition. For the standard Wilson action the
critical coupling is then given by $\beta = 5.0933(7)$ \cite{Alv92}.
The spatial geometry
is cylindrical. We have used the sizes
$L_x \times L_y \times L_z = 4 \times 4 \times 64$, $4 \times 6 \times 64$,
$6 \times 6 \times 64$, $6 \times 8 \times 96$
and $8 \times 8 \times 128$. In all cases $C$-periodic
boundary conditions have been used in the long $z$-direction. In the transverse
$x$- and $y$-directions we have used both periodic and $C$-periodic boundary
conditions. Of course, thermodynamics requires periodic boundary conditions
for the gluon field in the euclidean time direction. We have performed
measurements at several $\beta$-values very close to the critical coupling
using an overrelaxed heat bath algorithm.
To interpolate our results to other $\beta$-values we have used
standard reweighting techniques. Per lattice we have typically performed
50000-250000 measurements.

The low energy states in the spectrum of the transfer matrix in
the $z$-direction are extracted from the correlation functions of the operators
${\cal O}_i$ of eq.(\ref{operators}). Since we always use $C$-periodic
boundary conditions in the $z$-direction we introduce a complete set of
states $|n\rangle$ with their charge conjugate states $C|n\rangle$ and we
obtain
\begin{eqnarray}
Z \langle {\cal O}_i(0) {\cal O}_i(z) \rangle & = &
 \sum_n \langle n|{\cal O}_i \exp (-Hz) {\cal O}_i \exp \left(-H(L_z-z)\right)
C|n\rangle
 \nonumber\\
 & = & \sum_{m,n}
\langle n|{\cal O}_i |m\rangle\exp(-E_m z)\langle m|{\cal O}_i|n\rangle
                 \exp\left(-E_n(L_z-z)\right)c_n \nonumber\\
 & = & \sum_{m,n} c_n |\langle n|{\cal O}_i|m\rangle|^2
                 \exp(-E_m z) \exp\left(-E_n(L_z-z)\right)  , \nonumber \\
Z & = & \sum_n \langle n|\exp(- H L_z) C|n\rangle =
\sum_n c_n \exp(- E_n L_z),
\end{eqnarray}
where we have used $C|n\rangle = c_n |n\rangle$ for states of $C$-parity
$c_n = \pm 1$.

For $C$-periodic boundary conditions
in the $x$- and $y$-directions only the operators ${\cal O}_1$ and ${\cal O}_3$
are relevant. Both create the state with energy $E_1$. Putting $E_0 = 0$ and
denoting the states with energies $E_0$ and $E_1$ by $|0\rangle$ and
$|1\rangle$ the transfer matrix formalism predicts
\begin{eqnarray}
Z \langle {\cal O}_1(0) {\cal O}_1(z) \rangle & \sim &
|\langle 0|{\cal O}_1|1\rangle|^2(\exp(-E_1z) + \exp(-E_1(L_z-z))) \nonumber \\
& + & |\langle 0|{\cal O}_1|0\rangle|^2 +
|\langle 1|{\cal O}_1|1\rangle|^2 \exp(-E_1L_z), \nonumber \\
Z \langle {\cal O}_3(0) {\cal O}_3(z) \rangle & \sim &
|\langle 0|{\cal O}_3|1\rangle|^2(\exp(-E_1z) + \exp(-E_1(L_z-z))) \nonumber \\
& + & |\langle 0|{\cal O}_3|0\rangle|^2 +
|\langle 1|{\cal O}_3|1\rangle|^2 \exp(-E_1L_z), \nonumber \\
Z & \sim & 1 + \exp(-E_1L_z),
\end{eqnarray}
such that
\begin{eqnarray}
\langle {\cal O}_1(0) {\cal O}_1(z) \rangle & \sim &
A \cosh(E_1(z-\frac{1}{2}L_z)) + B, \nonumber \\
\langle {\cal O}_3(0) {\cal O}_3(z) \rangle & \sim &
C \cosh(E_1(z-\frac{1}{2}L_z)) + D.
\end{eqnarray}
We use these equations to determine the energy $E_1$ by fitting the
measured correlation functions.

When periodic boundary conditions are used in the $x$- and $y$-directions
the energy splittings $E_{1,2}-E_0$ and $E_3-E_0$ between the excited
states $|1\rangle,\,|2\rangle$ and $|3\rangle$ and the ground state
$|0\rangle$ become small close to the phase transition. Applying the
reduction rule
\begin{equation}
\begin{centering}
\begin{picture}(10,11.0)(0,-11)
\put(4,-3){\framebox(2,2)}
\put(6,-3){\framebox(2,2)}
\put(4,-5){\framebox(2,2)}
\put(10,-2){\makebox(0,0){$\;\times\;$}}
\end{picture}
\begin{picture}(10,11.0)(0,-11)
\put(4,-3){\framebox(2,2)}
\put(6,-3){\framebox(2,2)}
\put(4,-5){\framebox(2,2)}
\put(10,-2){\makebox(0,0){$\; = \;$}}
\end{picture}
\begin{picture}(12,11.0)(0,-11)
\put(4,-3){\framebox(2,2)}
\put(6,-3){\framebox(2,2)}
\put(8,-3){\framebox(2,2)}
\put(12,-2){\makebox(0,0){$\; + \;$}}
\end{picture}
\begin{picture}(10,11.0)(0,-11)
\put(4,-3){\framebox(2,2)}
\put(6,-3){\framebox(2,2)}
\put(4,-5){\framebox(2,2)}
\put(10,-2){\makebox(0,0){$\; + \;$}}
\end{picture}
\begin{picture}(10,11.0)(0,-11)
\put(4,-3){\framebox(2,2)}
\put(4,-5){\framebox(2,2)}
\put(4,-7){\framebox(2,2)}
\end{picture}
\end{centering}
\end{equation}
for $S_3$ (see \cite{Ham62}) and again setting $E_0 = 0$ we obtain
\begin{eqnarray}
Z \langle {\cal O}_1(0) {\cal O}_1(z) \rangle & \sim &
|\langle 0|{\cal O}_1|1\rangle|^2(\exp(-E_1z) + \exp(-E_1(L_z-z))) \nonumber \\
& + & |\langle 1|{\cal O}_1|3\rangle|^2 \exp(-E_1L_z) \nonumber \\
& \times & (\exp(-(E_3-E_1)z) + \exp(-(E_3-E_1)(L_z-z))), \nonumber \\
Z \langle {\cal O}_2(0) {\cal O}_2(z) \rangle & \sim &
|\langle 0|{\cal O}_2|2\rangle|^2(\exp(-E_2z) - \exp(-E_2(L_z-z))) \nonumber \\
& - & |\langle 2|{\cal O}_2|3\rangle|^2 \exp(-E_2L_z) \nonumber \\
& \times & (\exp(-(E_3-E_2)z) - \exp(-(E_3-E_2)(L_z-z))), \nonumber \\
Z \langle {\cal O}_3(0) {\cal O}_3(z) \rangle & \sim &
|\langle 0|{\cal O}_3|3\rangle|^2(\exp(-E_3z) + \exp(-E_3(L_z-z))) \nonumber \\
& + & |\langle 0|{\cal O}_3|0\rangle|^2 +
|\langle 3|{\cal O}_3|3\rangle|^2 \exp(-E_3L_z), \nonumber \\
Z & \sim & 1 + \exp(-E_3L_z).
\label{correlations}
\end{eqnarray}
Here the additional symmetry relation
$| \langle 1|{\cal O}_i |1 \rangle |^2 = | \langle 2| {\cal O}_i |2\rangle |^2$
was used.
Note that the correlation function of the $C$-odd operator ${\cal O}_2$ is
anti-periodic because of $C$-periodic boundary conditions in the $z$-direction.
The $S_3$-symmetry ensures that $E_1 = E_2$,
$|\langle 0|{\cal O}_1|1 \rangle|^2 =
|\langle 0|{\cal O}_2|2 \rangle|^2$ and
$|\langle 1|{\cal O}_1|3 \rangle|^2 =
|\langle 2|{\cal O}_2|3 \rangle|^2$. For very large $L_z$ such that
$E_iL_z \gg 1$ eq.(\ref{correlations}) simplifies to
\begin{eqnarray}
\langle {\cal O}_1(0) {\cal O}_1(z) \rangle & \sim &
A \cosh(E_1(z-\frac{1}{2}L_z)), \nonumber \\
\langle {\cal O}_2(0) {\cal O}_2(z) \rangle & \sim &
B \sinh(E_2(z-\frac{1}{2}L_z)), \nonumber \\
\langle {\cal O}_3(0) {\cal O}_3(z) \rangle & \sim &
C \cosh(E_3(z-\frac{1}{2}L_z)) + D.   \label{fitper}
\end{eqnarray}
In most of our fits we have used this simpler form. However,
in some cases we have checked
our results using the more general expressions (\ref{correlations}) and we
found agreement within the error bars.

Fig.1 shows a configuration on the
$8 \times 8 \times 128 \times 2$ lattice which consists of several
bulk phases separated by interfaces.
%
\begin{figure}
\vspace*{8.5cm}
\caption{ {
The $z$-dependence of ${\cal O}_1(z)=
\sum_{x,y} \Phi_1(x,y,z)$ for a
configuration on an 8$\times$8$\times$128$\times$2 lattice
with $C$-periodic boundary conditions in the spatial directions
at $\beta=5.0933$.
}}
\label{fig1}
\end{figure}
Two typical correlation
functions are depicted in fig.2.
\begin{figure}
\vspace*{10cm}
\caption{ {
The correlation functions
$\langle {\cal O}_1(0) {\cal O}_1(z) \rangle $ and
$\langle {\cal O}_2(0) {\cal O}_2(z) \rangle $
on a $4\times 6 \times 64 \times 2$ lattice with
$C$-periodic boundary condition only in the $z$-direction.
The solid lines represent fits of the data to the corresponding
equation (\protect{\ref{fitper}}).
}}
\label{fig2}
\end{figure}
We should mention that autocorrelation times
of the correlation functions are rather long,
especially at large distances. Therefore, it was always a good check if
the energies $E_1$ and $E_2$ agreed within errors (as they should because of
$S_3$-symmetry).


Fig.3a
shows the energy $E_1$ on a $4 \times 6 \times 64 \times 2$ lattice
with $C$-periodic boundary conditions in the $x$- and $y$-directions.
\begin{figure}
\vspace*{14cm}
\caption{ {
The spectra of the lowest energies in the region of the
phase transition temperature.
In a) the energy $E_{1}$ from simulations on a
$4\times 6 \times 64 \times 2$ lattice with
$C$-periodic boundary conditions in the $x$- and $y$-direction is shown.
In b) the degenerate energy $E_{1,2}$ and the
energy $E_3$ derived from simulations on a
$6\times 6 \times 64 \times 2$ lattice with
periodic boundary conditions in the $x$- and $y$-direction are shown.
The energies are plotted as boxes at the  simulation points,
whereas the interpolated values are plotted as diamonds.
}}
\label{fig3}
\end{figure}
As expected, there is a minimum at the critical coupling. From the finite
size effect of the value at the minimum we have determined
the reduced confined-deconfined interface tension using eq.(\ref{finiteC}).
The result of the fit, which is depicted in fig.4,
\begin{figure}
\vspace*{11cm}
\caption{ {
The dependence of the energies on the cross section
$A=L_x L_y$ of the cylindrical lattices.
The values of $\log(E_1)$ are shown
from simulations on lattices with $C$-periodic boundary conditions
in the $x$- and $y$-direction (diamonds) and
the values of $\log (E_{1,2})$ ($=\log(\frac{1}{2}E_3)$ ) are shown
from simulations on lattices with periodic boundary conditions
in the $x$- and $y$-direction (squares). The solid lines represent
the fits of the data to eq.(\protect{\ref{finiteC}})
and eq.(\protect{\ref{finiteP}}) respectively.
}}
\label{fig4}
\end{figure}
is given by
\begin{equation}
\sigma_{cd} = 0.035(1) = 0.140(4) T_c^2 \,\,\,\,\,\,\, \mbox{($C$-periodic)}.
\end{equation}
The fitted value of the pre-exponential factor is $\delta = 0.196(5)$
which is much
smaller than the prediction $\delta = 1$ for rigid interfaces \cite{Bor91}.
We take this as an indication that the confined-deconfined interface is indeed
rough.

Fig.3b shows the energies $E_{1,2}$ and $E_3$ on a
$6 \times 6 \times 64 \times 2$ lattice with periodic boundary conditions
in the $x$- and $y$-directions. As expected, the energy $E_3$ has a minimum
which is, however, slightly shifted away from the critical $\beta$ into the
deconfined phase. At the same time, the point where $2 E_{1,2} = E_3$ is
shifted slightly into the confined phase. In case of incomplete wetting one
would expect such shifts, but they would go exactly in the opposite direction.
Therefore, our data favour complete wetting. Of course,
then eq.(\ref{spectrum}) predicts that the minimum of $E_3$ and the point where
$2 E_{1,2} = E_3$ should correspond to the same $\beta$-value, which is
almost --- but not quite --- the case for our data. We believe that this effect
is some sub-leading finite size effect which is not correctly reproduced by
our dilute interface calculation. The same effect is visible in the data
of Wang and DeTar \cite{Wan92} in the three-dimensional
three state Potts model, which also shows
complete wetting \cite{Kar91}. There the shift is
present on smaller lattices, and it disappears on larger volumes,
supporting the assumption of a sub-leading effect. Altogether, we consider
fig.3b as a strong indication that the deconfined-deconfined interfaces
are
completely wet by the confined phase at the deconfinement phase transition.
To extract the reduced confined-deconfined interface tension from our periodic
lattices we have taken the value of $E_{1,2}$ at the point where $2 E_{1,2} =
E_3$ and we have fitted its finite size behavior to eq.(\ref{finiteP}). The
fit, which is also shown in fig.4, results in
\begin{equation}
\sigma_{cd} = 0.034(2) = 0.136(8) T_c^2 \,\,\,\,\,\,\, \mbox{(periodic)}.
\end{equation}
It is reassuring (and further supports the assumption of complete wetting)
that this result is in agreement with the one obtained from
the $C$-periodic simulation.
Also the fitted value of the pre-exponential factor,
$\delta = 0.202(6)$, agrees with the one obtained from the
$C$-periodic data. This value again
indicates that the confined-deconfined
interface is rough.
Furthermore, it means that the capillary waves have the
same behavior in periodic and in $C$-periodic volumes. This was expected
because they are small fluctuations which are insensitive to global
topological characteristics of the boundary conditions. Taking the results
of both boundary conditions together we state as our final
answer for the reduced confined-deconfined interface tension
\begin{equation}
\sigma_{cd} = 0.139(4) T_c^2.
\end{equation}

We have also re-analyzed the data of Wang and DeTar \cite{Wan92}
from our perspective. They have measured the energies $E_{1,2}$ and $E_3$ in
the three-dimensional
three state Potts model using periodic boundary conditions. As Wang and
DeTar have pointed out, their results are in good agreement with
complete wetting.
On a $20^2 \times 120$ lattice they
find $E_{1,2} = 0.048(1)$ at $\beta = 0.3668$ and
$E_{1,2} = 0.0397(6)$ at $\beta = 0.3669$
while at $\beta = 0.3670$ on a $30^2 \times 120$
lattice $E_{1,2} = 0.0185(3)$. In all cases $2 E_{1,2} = E_3$ within the
error bars. Assuming that the asymptotic formulae (\ref{finiteP}) are valid
for these lattices we obtain
\begin{equation}
\sigma_{cd} = 0.0017(3) \,\,\,\,\,\,\, \mbox{(Potts model)},
\label{Potts}
\end{equation}
where the error is mostly systematical. This is due to the same slight
shift of the minimum of $E_3$ away from the point where $2 E_{1,2} = E_3$
which we also observed in our data. In the Potts model the shift is present
only
on the smaller lattice, while it disappears on the larger volume. Perhaps the
result of eq.(\ref{Potts}) should be taken with a grain of salt because
it relies only on two different volumes. For the pre-exponential factor
we find $\delta = 0.042(4)$ indicating that the ordered-disordered interface
is rough. The ordered-ordered reduced interface
tension $\sigma_{dd}(T_c)$ has been determined in the Potts model
by Karsch and Patk\'{o}s \cite{Kar91}. They enforced the interface by boundary
conditions and computed the free energy difference due to the interface
by a numerical integration. Their result
\begin{equation}
\sigma_{dd}(T_c) \geq 0.0026(2) \,\,\,\,\,\,\, \mbox{(Potts model)}
\end{equation}
(translated to our notation)
is consistent with eq.(\ref{Potts}) because complete wetting implies
$\sigma_{dd}(T_c) = 2 \sigma_{cd}$.

\section{Conclusions}

We have related the finite size effects of the spectrum of the transfer matrix
to the value of the reduced confined-deconfined interface tension both for
$C$-periodic and for periodic boundary conditions. In the periodic case one
must distinguish between the possibilities of complete and incomplete wetting.
Our numerical data strongly suggest that the deconfined-deconfined domain
walls of the $SU(3)$ pure gauge theory are completely wet by the confined
phase.
Hence, in a hypothetical universe, which contains only gluons but no quarks
(or other particles), the deconfinement phase transition would proceed via
domain wall splitting and not via the usual bubble nucleation. However,
the presence of quarks in our universe made wetting at the
deconfinement phase transition impossible \cite{Jue92,Ign92}. Still, the
reduced
confined-deconfined interface tension of the pure glue system serves as
an upper limit on the corresponding value in real QCD. The determination
of the value $\sigma_{cd} = 0.139(4) T_c^2$
required a careful analysis of finite size
effects and an understanding of the wetting dynamics.

We like to mention that the transfer matrix method is a very powerful tool
to extract interface properties from numerical simulations, especially
because finite size effects are well understood. Certainly, in this method
information on correlation functions at large distances is needed. With only
local algorithms available (as it the case for gauge theories) this is not
an easy task, and it requires very high statistics. An attractive alternative
is the multicanonical algorithm of Berg and Neuhaus \cite{Ber92}. We have also
applied this algorithm to the pure $SU(3)$ gauge theory \cite{Gro92} and
we have measured the interfacial free energy. It has turned out that the
use of cylindrical lattices is again advantageous in order to reduce the
interactions between interfaces. This work is presently in progress.

\section*{Acknowledgement}

We like to thank C. Borgs and V.~Privman for interesting
discussions about
 finite size effects at first order phase transitions.
One of us (T.T.) would like to thank J. Jers\'{a}k for
discussions.
Our numerical simulations were performed on
the CRAY Y-MP at the HLRZ in J\"ulich and on the NEC SX-3 at the University of
K\"oln.

\section*{Appendix}
\renewcommand{\theequation}{A.\arabic{equation}}

In this appendix we would like to give some of the details for
the formulae occurring in section 3. We shall start with the
{\bf complete wetting} case.
Let us also assume that we have $q$ deconfined
phases, such that $q=3$ for periodic and $q=1$ for
$C$-periodic boundary conditions.
It is convenient to define the following integrals:
\begin{eqnarray}
I_{n}(y,\xi) & = & z^{-n} \int_{\xi}^{z} d\xi_{1}
                      \int_{\xi_{1} }^{z} d\xi_{2}
             \cdots \int_{\xi_{n-1} }^{z} d\xi_{n}
\exp[y \sum_{i=1}^{n}(-1)^{i}\xi_{i} ] \nonumber \\
             & = & \int_{\xi}^{1} d\xi_{1} \int_{\xi_{1} }^{1} d\xi_{2}
             \cdots \int_{\xi_{n-1} }^{1} d\xi_{n}
\exp[zy \sum_{i=1}^{n}(-1)^{i}\xi_{i} ], \;\;\;n\geq 1
\end{eqnarray}
where $I_{n}(y,1) = 0$ and $y = 2x = (f_{c} - f_{d})A/T$.
The aim is to evaluate
$I_{n}(y) \equiv I(y,0)$, which exactly corresponds
to $n$ confined-deconfined interfaces. For reasons of notational simplicity
we rescale $z x \rightarrow x$, $z y \rightarrow y$, and
$z\sqrt{q} \delta \exp(-\sigma_{cd}A) = z \delta' \rightarrow \delta'$.
The different interface contributions $\sqrt{q}t_{cd},t_{cc}$
and $t_{dd}+(q-1)t_{dd'}$
can now be expressed in terms of
$I_{n}(y)$:
\begin{eqnarray}
\sqrt{q}t_{cd} & = &
       \exp(x) \sum_{n=0}^{\infty}I_{2n+1}(y)\delta'^{2n+1}, \nonumber \\
t_{cc} & = &
           \exp(-x) (1+\sum_{n=1}^{\infty}I_{2n}(y)\delta'^{2n}), \nonumber \\
t_{dd} + (q-1)t_{dd'}& = &
           \exp(x) (1+\sum_{n=1}^{\infty}I_{2n}(-y)\delta'^{2n}).
\label{A4}
\end{eqnarray}
 The case $n = 1$ is trivial and yields
\begin{eqnarray}
  I_{1}(y,\xi) & = & \frac{1}{y}(\exp(-y\xi) - \exp(-y)), \nonumber \\
  I_{1}(y) & = & \frac{1}{y}(1 - \exp(-y)).
\end{eqnarray}
For $n\geq 2$ we find after
differentiating $I_{n}(y,\xi)$     with respect to $\xi$
\begin{equation}
  \frac{\partial I_{n}(y,\xi)}{\partial \xi} =
     - \exp(-y\xi)I_{n-1}(-y,\xi).
\end{equation}
In particular for $n=2$:
\begin{equation}
  \frac{\partial I_{2}(y,\xi)}{\partial \xi} =
                 \frac{1}{y} (1 - \exp[y(1-\xi)]).
\end{equation}
The integration is easy and we obtain
\begin{eqnarray}
I_{2}(y,\xi) & = &\frac{\xi - 1}{y} + \frac{1}{y^2}(\exp[y(1-\xi)] - 1),
\nonumber \\
I_{2}(y) & = & - \frac{1}{y} + \frac{1}{y^2}(\exp(y) - 1)
 = - \exp(y)\frac{\partial I_{1}(y)}{\partial y}.
\end{eqnarray}
Repeating the steps above we arrive at
\begin{eqnarray}
  I_{2n}(y) & = & - \frac{1}{n}
               \exp(y)\frac{\partial I_{2n-1}(y)}{\partial y}, \nonumber  \\
  I_{2n+1}(y) & = &  \frac{1}{n}
               \exp(-y)\frac{\partial I_{2n}(y)}{\partial y},
               \;\;\; n\geq 1.
\label{A13}
\end{eqnarray}
This we checked explicitly up to $n = 3$. From the previous two
equations follows:
\begin{eqnarray}
  I_{2n+1}(y) & = & \frac{(-1)^n}{n!^2}
               \left[\frac{\partial   }{\partial y} +
                \frac{\partial ^2}{\partial y^2}\right]^{n}I_{1}(y),
\nonumber \\
  I_{2n+2}(y) & = & \frac{(-1)^n}{n!(n+1)!}
               \left[- \frac{\partial   }{\partial y} +
                  \frac{\partial ^2}{\partial y^2}\right]^{n}I_{2}(y),
               \;\;\; n\geq 1.
\end{eqnarray}
Now we can do the $\sqrt{q}t_{cd}$ term. Using the identities:
\begin{equation}
       \exp(x)I_{2n+1}(2x) = \frac{4^n}{n!^2}   \sum_{i=0}^{n}
               \left(
               \begin{array}{c}
                  n \\ i
               \end{array}
               \right)
 (-1)^{i}\frac{\partial ^{2i}}{\partial x^{2i}} (\exp(x)I_{1}(2x))
                             = \frac{1}{n!}
         \frac{\partial ^{n}}{(\partial x^{2})^n}
         \left(\frac{\sinh x}{x}\right),
\end{equation}
we find
\begin{equation}
\sqrt{q}t_{cd} = \sum_{n=0}^{\infty}\frac{1}{n!}
         \frac{\partial ^{n}}{(\partial x^{2})^n}
         \left(\frac{\sinh x}{x}\right)\delta'^{2n+1} \\
          = \frac{\delta'}{\sqrt{x^{2} + \delta'^2}}
               \sinh \sqrt{x^{2} + \delta'^2}.
\label{A19}
\end{equation}
Now we come to $t_{cc}$. Using eqs. (\ref{A4}) and (\ref{A13}) we can write:
\begin{equation}
t_{cc} = \exp(-x) - \exp(x)
\frac{\partial}{\partial x}
\sum_{n=1}^{\infty}\frac{1}{2n}I_{2n-1}(y)\delta'^{2n}.
\end{equation}
Differentiating the previous equation
with respect to $\delta'$ and integrating  we obtain
\begin{eqnarray}
t_{cc} & = &
\exp(-x) - \exp(x)\int_{0}^{\delta'} d\delta'
\frac{\partial}{\partial x}(\exp(-x)\sqrt{q}t_{cd}) \nonumber \\
            & = &
               \cosh \sqrt{x^{2} + \delta'^2} -
            \frac{x}{\sqrt{x^{2} + \delta'^2}}
               \sinh \sqrt{x^{2} + \delta'^2}.
\end{eqnarray}
Likewise we get
\begin{equation}
t_{dd} + (q-1)t_{dd'} =
               \cosh \sqrt{x^{2} + \delta'^2} +
            \frac{x}{\sqrt{x^{2} + \delta'^2}}
               \sinh \sqrt{x^{2} + \delta'^2}.
\end{equation}
Finally we have trivially
\begin{equation}
   t_{dd} - t_{dd'} = \exp(x).
\end{equation}

The case of {\bf incomplete wetting} is somewhat harder. First we
rescale the expression
$z(q-1) \gamma \exp(-\sigma_{dd}A) =$ $z \gamma' \rightarrow \gamma'$.
We begin again with $\sqrt{q}t_{cd}$.
To lowest non-trivial order in $\delta'$, but to all orders in $\gamma'$,
we obtain easily
\begin{equation}
\sqrt{q}t_{cd} = \exp(-x)\sum_{n=0}^{\infty} \frac{1}{n!}
         \frac{\partial ^n}{\partial y^n}I_{1}(-y) \gamma'^{n}  \\
        = \frac{\exp(x)}{y}\sum_{n=0}^{\infty}
       \left( -\frac{\gamma'}{y} \right)^{n}
    \left( \sum_{i=0}^{n}\frac{(-y)^i}{i!} - \exp(-y)\right).
\end{equation}
The single sum is easy
\begin{equation}
  \frac{1}{y}\sum_{n=0}^{\infty}
  \left(-\frac{\gamma'}{y}\right)^{n}
            = \frac{1}{y+\gamma'}.
\end{equation}
For the double sum we first let $i \rightarrow n-i$ and then we
interchange the order of the summations. This gives
\begin{equation}
  \frac{1}{y}\sum_{n=0}^{\infty}
  \left(-\frac{\gamma'}{y}\right)^{n}
             \sum_{i=0}^{n}\frac{(-y)^{n-i}}{(n-i)!}
              = \frac{1}{y}\sum_{i=0}^{\infty}
  \left(-\frac{\gamma'}{y}\right)^{i}
             \sum_{n=i}^{\infty}\frac{\gamma'^{n-i}}{(n-i)!} \\
              = \frac{\exp(\gamma')}{y+\gamma'},
\label{A29}
\end{equation}
so that with $x' = x + \gamma'/2$
\begin{equation}
  \sqrt{q}t_{cd} = \delta' \frac{\sinh x'}{x'}\exp(\gamma'/2).
\end{equation}
Combining eqs. (\ref{A19}) and (\ref{A29}) we conjecture that to all orders
in $\delta'$ the following holds:
\begin{equation}
 \sqrt{q}t_{cd} = \frac{\delta'}{\sqrt{x'^{2} + \delta'^2}}
           \sinh \sqrt{x'^{2} + \delta'^2}\exp(\gamma'/2).
\end{equation}
This we checked explicitly up to order $O(\delta'^4)$. It is now easy
to guess the expressions for $t_{cc}$ and $t_{dd} + (q-1)t_{dd'}$. They
are:
\begin{eqnarray}
  t_{cc} & = & \left(\cosh \sqrt{x'^{2} + \delta'^2} -
         \frac{x'}{\sqrt{x'^{2} + \delta'^2}}
                     \sinh \sqrt{x'^{2} + \delta'^2}\right)
                    \exp(\gamma'/2), \nonumber \\
 t_{dd} + (q-1)t_{dd'} & = &
              \left( \cosh \sqrt{x'^{2} + \delta'^2} +
         \frac{x'}{\sqrt{x'^{2} + \delta'^2}}
                     \sinh \sqrt{x'^{2} + \delta'^2} \right)
                    \exp(\gamma'/2).
\end{eqnarray}
This we also checked explicitly up to order $O(\delta'^4)$.
Notice for $\gamma' = 0$ we reproduce the results for the complete
wetting case.
Finally we find
\begin{equation}
  t_{dd} - t_{dd'} = \exp(x)\exp[-\gamma'/(q-1)].
\end{equation}
The formulae given in the main text are obtained by replacing
\begin{equation}
   x \rightarrow zx, \; x' \rightarrow  zx', \; \delta' \rightarrow z \delta'=
z \sqrt{q} \delta \exp(- \sigma_{cd}A), \; \gamma' \rightarrow z \gamma'=
z(q-1) \gamma \exp(-\sigma_{dd}A).
\end{equation}
As an addendum we would like to mention the following identity:
\begin{equation}
 \mbox{det}(t) = \exp[-(f_{c}+qf_{d})A/T],
\end{equation}
which is equivalent to
\begin{equation}
  t_{cc}(t_{dd}+(q-1)t_{dd'}) - \sqrt{q}t_{cd}\sqrt{q}t_{cd} =
  \exp(\gamma').
\end{equation}

\end{document}